\documentclass[jcp,aip,10pt,floatfix,twocolumn,showpacs]{revtex4-1} 
\usepackage{color,soul}
\usepackage{graphicx}
\usepackage{siunitx}
\usepackage[breaklinks=true,colorlinks=true,linkcolor=blue,urlcolor=blue,citecolor=blue]{hyperref}

\usepackage{dblfloatfix}
\usepackage{amsmath, amsthm, amssymb}
\usepackage{amsthm}
\usepackage{multirow}
\usepackage[normalem]{ulem}
\usepackage{soul}
\usepackage{makecell}
\usepackage[table]{xcolor}
\definecolor{Gray}{gray}{0.85}
\definecolor{LightCyan}{rgb}{0.88, 1, 1}
\definecolor{Apricot}{rgb}{0.98, 0.81, 0.69}
\usepackage{threeparttable}
\newcommand{\be}{\begin{equation}}
\newcommand{\ee}{\end{equation}}
\newcommand{\bea}{\begin{eqnarray}}
\newcommand{\eea}{\end{eqnarray}}

\usepackage[rightcaption,]{sidecap}
\sidecaptionvpos{figure}{c}
\usepackage[utf8]{inputenc}
\usepackage{amsmath}
\usepackage{amssymb}
\usepackage{graphicx}
\usepackage{hyperref}
\usepackage[export]{adjustbox}
\usepackage[update]{epstopdf}
\usepackage{multirow}

\begin{document}

\title{Bridging-induced Aggregation in Neutral Polymers: Dynamics and Morphologies}
\author{Hitesh Garg}
\email{hiteshgarg@imsc.res.in}
\affiliation{The Institute of Mathematical Sciences, C.I.T. Campus,
Taramani, Chennai 600113, India}
\affiliation{Homi Bhabha National Institute, Training School Complex, Anushakti Nagar, Mumbai, 400094, India}

\author{Satyavani Vemparala}
\email{vani@imsc.res.in}
\affiliation{The Institute of Mathematical Sciences, C.I.T. Campus,
Taramani, Chennai 600113, India}
\affiliation{Homi Bhabha National Institute, Training School Complex, Anushakti Nagar, Mumbai, 400094, India}

\begin{abstract}
Using molecular dynamics simulations, we investigate the aggregation behavior of neutral stiff (rod-like) and flexible polymer chains mediated by attractive crowders. Attractive crowders serve as bridging agents, inducing aggregation through effective intra-polymer attractions. The critical monomer-crowder attraction strength ($\epsilon_{mc}^*$) required for aggregation differs notably between rigid rods and flexible polymers. Interestingly, this aggregation threshold closely matches the critical attraction required for the extended-to-collapsed (coil-globule) transition of a single flexible polymer chain, suggesting a fundamental connection between single-chain collapse and multi-chain aggregation. Furthermore, we demonstrate that $\epsilon_{mc}^*$ decreases with increasing system density and larger crowder sizes, highlighting the synergistic roles of crowding effects and crowder dimensions. Aggregate morphologies exhibit strong dependence on polymer flexibility: rigid rods predominantly form elongated cylindrical bundles, whereas flexible polymers aggregate into compact spherical clusters. These findings provide comprehensive insights into how bridging interactions driven by attractive crowders regulate polymer aggregation dynamics and morphologies, emphasizing the importance of polymer rigidity, crowder size, and system density.
\end{abstract}

\maketitle
\section{Introduction}

Non-covalent interaction-driven self-assembly, phase separation, and polymer aggregation are central phenomena influencing numerous biological processes and material design strategies~\cite{whitesides2002beyond,sanchez2021aggregation,rajagopal2004self}. These interactions underpin the formation of organized structures with specific mechanical and functional properties, making polymer aggregation a critical focus area in understanding cellular organization and developing tailored synthetic materials~\cite{walheim1999morphologies,Brown1997-rx,Braun2002-de,Whaley2000-va,Goede2004-py,bachmann2010microscopic}. Aggregation in biological systems exhibits dual roles, serving essential structural and functional purposes or contributing to pathological conditions. Beneficial examples include cytoskeletal organization driven by the aggregation of rod-like actin filaments and microtubules, and the formation of biomolecular condensates via liquid-liquid phase separation (LLPS), enabled by weak, multivalent interactions among proteins and nucleic acids~\cite{graham2023liquid,castaneda2021regulation,ray2020alpha}. Conversely, pathological aggregation underlies severe diseases, including Alzheimer’s and Parkinson’s, where amyloid-beta and alpha-synuclein aggregation lead to neurotoxic assemblies, or thrombosis, characterized by uncontrolled platelet aggregation and severe vascular complications~\cite{webber2020pathophysiology,mukherjee2023liquid}. Aggregation processes in biological systems are typically influenced by molecular crowding, where crowding agents interact with polymers through repulsive or attractive forces, introducing additional complexity to the aggregation dynamics~\cite{porter2019interplay,galanis2010depletion,hu2013depletion,speer2022macromolecular,shah2011effects}.

In synthetic and engineered systems, polymer aggregation is equally crucial, particularly in stabilizing multi-phase mixtures such as emulsions, where controlled aggregation prevents undesirable phase separation and ensures long-term functionality~\cite{cates2008bijels,bandyopadhyay2006slow,alison2016pickering,gibaud2012new}. In these contexts, achieving arrested coarsening—a state where aggregates reach a stable, finite size—is desirable in industries ranging from pharmaceuticals and food to cosmetics~\cite{pawar2011arrested,gao2015microdynamics}. Approaches to achieve arrested coarsening include modification of interfacial tension with surfactants or polymers, particle jamming, and notably Pickering emulsions, where solid particles irreversibly adsorb onto droplet surfaces, creating stable physical barriers against coalescence~\cite{pickering1907cxcvi,ramsden1904separation,chevalier2013emulsions,tcholakova2008comparison}. Among stabilizing agents, anisotropic particles such as rod-shaped polymers and natural materials like cellulose and chitosan exhibit unique advantages by forming interconnected networks, reducing coalescence effectively at low concentrations~\cite{hunter2020pickering,coertjens2017adsorption,dugyala2016nano,kalashnikova2011new}. Polymer aggregation can generally occur through two distinct mechanisms: \textit{depletion interactions} and \textit{bridging interactions}. Depletion interactions occur when non-adsorbing crowders create osmotic pressure differences that push polymer segments closer, thereby compacting the polymers~\cite{asakura1954interaction,yodh2001entropically}. Such interactions depend critically on crowder size and polymer conformational flexibility, with larger crowders typically exerting stronger depletion effects~\cite{kim2015polymer}. Conversely, bridging interactions arise when crowders bind simultaneously to two polymer segments, physically linking them and promoting stable aggregation~\cite{heyda2013rationalizing,sapir2016macromolecular}. Bridging interactions can thus result in distinct aggregation kinetics and morphologies compared to purely depletion-driven aggregation, offering unique pathways to control polymer assembly.

Neutral polymer systems—characterized by inherently repulsive monomer-monomer interactions—provide an ideal model to explore bridging-driven aggregation. Attractive crowders in such systems can mediate polymer-polymer interactions through bridging, effectively creating intra- and inter-polymer attractions~\cite{antypov2008computer,rodriguez2015mechanism,sagle2009investigating,brackley2020polymer,garg2023conformational}. Weak polymer-crowder attraction encourages extended conformations, maximizing crowder interactions, whereas stronger attractions convert crowders into bridging agents or "stickers", inducing collapsed configurations through direct monomer-crowder-monomer links. This behavior mirrors counterion-driven aggregation observed in charged polymer systems, where multivalent counterions bridge charged segments, promoting compact aggregates~\cite{Tripathi2023-wu,varghese2012aggregation,tom2016aggregation,tom2017aggregation}. Bridging interactions specifically are relevant in biological contexts, where molecular crowding influences cytoskeletal dynamics and biomolecular condensate formation, phenomena similarly observed in engineered polymer systems, underscoring their broad importance~\cite{wang2013phase,anderson2002insights,rajan2021review,wang2018protein}. Recent experiments show that polymers can aggregate cells or macromolecules either by adsorbing onto them (bridging) or by exclusion volume effects (depletion)~\cite{secor2018entropically}. Whereas depletion-induced phase separation in polymer solutions has been extensively studied, the collective behavior of multiple polymers under the influence of attractive crowders (bridging) remains comparatively unexplored.

In this study, we specifically investigate aggregation driven by attractive bridging crowders in neutral polymer systems. Using molecular dynamics simulations, we quantify the critical polymer-crowder attraction strength required for aggregation, examining how this threshold depends on polymer flexibility, crowder size, and system density. Further, we evaluate dynamical scaling exponents, providing mechanistic insights into aggregation dynamics. Interestingly, we find that the minimum crowder attraction needed to aggregate flexible polymers coincides with the attraction needed to collapse a single polymer chain – suggesting a fundamental link between single-chain compaction and multi-chain aggregation. Beyond equilibrium configurations, we analyze the aggregation dynamics via scaling exponents and characterize the resulting cluster geometries. These results offer fundamental understanding of bridging-driven aggregation processes, informing rational design in both biological and synthetic polymer contexts.

\section{Model and Methods}\label{Sec-2}
We consider a system consisting of $N_p = 300$ neutral polymer chains, each composed of $N_m = 30$ monomers, and $N_c = 9000$ neutral spherical crowders confined in a cubic simulation box of side length $L$. The polymers and crowders are initially arranged periodically and equilibrated to achieve a homogeneous distribution. Simulations are conducted at varying densities by adjusting the box size ($L = 100, 150, 200, 250, 300$, in units of $\sigma_m$), corresponding to crowder number densities $\phi_c = N_c/V$ of $9 \times 10^{-3}, 2.66 \times 10^{-3}, 1.12 \times 10^{-3}, 5.76 \times 10^{-4},$ and $3.33 \times 10^{-4}$. Crowder sizes are generally equal to the monomer size ($\sigma_c = \sigma_m$), but additional simulations explore the effect of crowder size with $\sigma_c = 0.5, 1.0, 2.0$ while maintaining constant volume density.

Adjacent monomers within a polymer chain are connected by harmonic springs: \begin{equation}
 V_{bond}(r)=\frac{1}{2} k (r-b)^2, 
 \end{equation} 
 where $b = 1.12\sigma_m$ is the equilibrium bond length, and $k = \frac{500\epsilon_{mm}}{\sigma_{mm}^2}$ is the stiffness constant. Polymer rigidity is controlled by a three-body angular potential: 
\begin{equation} 
V_{angle}(\theta)= k_{\theta}[1+\cos\theta], 
\end{equation} 
where $\theta$ is the angle between adjacent bonds. To model varying chain flexibilities, $k_{\theta}$ is assigned a value of $0$ for fully flexible case and $10^3$ for the fully rigid case.

Non-bonded interactions between particles are governed by the Lennard-Jones (LJ) potential: 
\begin{equation} 
V_{LJ} (r)= 4\epsilon_{ij}\left[ \left(\frac{\sigma_{ij}}{r} \right)^{12}-\left(\frac{\sigma_{ij}}{r} \right)^6 \right];  r \le r_c 
\end{equation} 
where $i,j$ refer to monomers ($m$) or crowders ($c$). In the text, we denote the size of the monomer and crowder as $\sigma_m$ and $\sigma_c$ respectively. The cutoff distance $r_c$ is set to $1.12\sigma_m$ for monomer-monomer interactions to model purely repulsive Weeks-Chandler-Andersen (WCA) interactions, and $3.00\sigma_m$ for monomer-crowder and crowder-crowder interactions to incorporate attractive forces. For all simulations, the crowder-crowder interaction strength is fixed at $\epsilon_{cc} = 1.0$, while the monomer-crowder attraction strength $\epsilon_{mc}$ is varied between $0.1$ and $4.0$.

Simulations are performed using the MD LAMMPS software package~\cite{plimpton1995fast}, with visualization of trajectories and snapshots carried out using the Visual Molecular Dynamics (VMD) package~\cite{HUMP96}. The equations of motion are integrated using the velocity-Verlet algorithm with a time step $\delta t = 0.001\tau$, where $\tau = \sigma_{m}\sqrt{m/\epsilon_{mm}}$ represents the simulation time unit. All simulations are conducted under constant temperature ($T = 1.0$) using a Nose-Hoover thermostat. The system is initially equilibrated in the $NVT$ ensemble, followed by $NPT$ equilibration to adjust the box volume to the desired density. A final equilibration in the $NVT$ ensemble ensures homogeneous mixing, as illustrated in Figure~S1. The simulations are run for $2 \times 10^7$ steps to achieve statistical equilibrium.

Block averaging is performed over the last $n = 900$ frames of each simulation, with each block containing 100 frames, resulting in $n_b = 9$ blocks. The standard deviation across blocks is divided by $\sqrt{n_b}$ to calculate errors. Additionally, three independent simulations with different random seeds for initial velocities are conducted for each parameter set to ensure robustness. Aggregate sizes and other observables are calculated using averaged configurations from the equilibrated trajectories.

\section{\label{sec:results} Results}
\subsection{\label{sec:kinetics} Aggregation of rigid and flexible polymers in the presence of bridging crowders}
\begin{figure}
\includegraphics[width=\columnwidth]{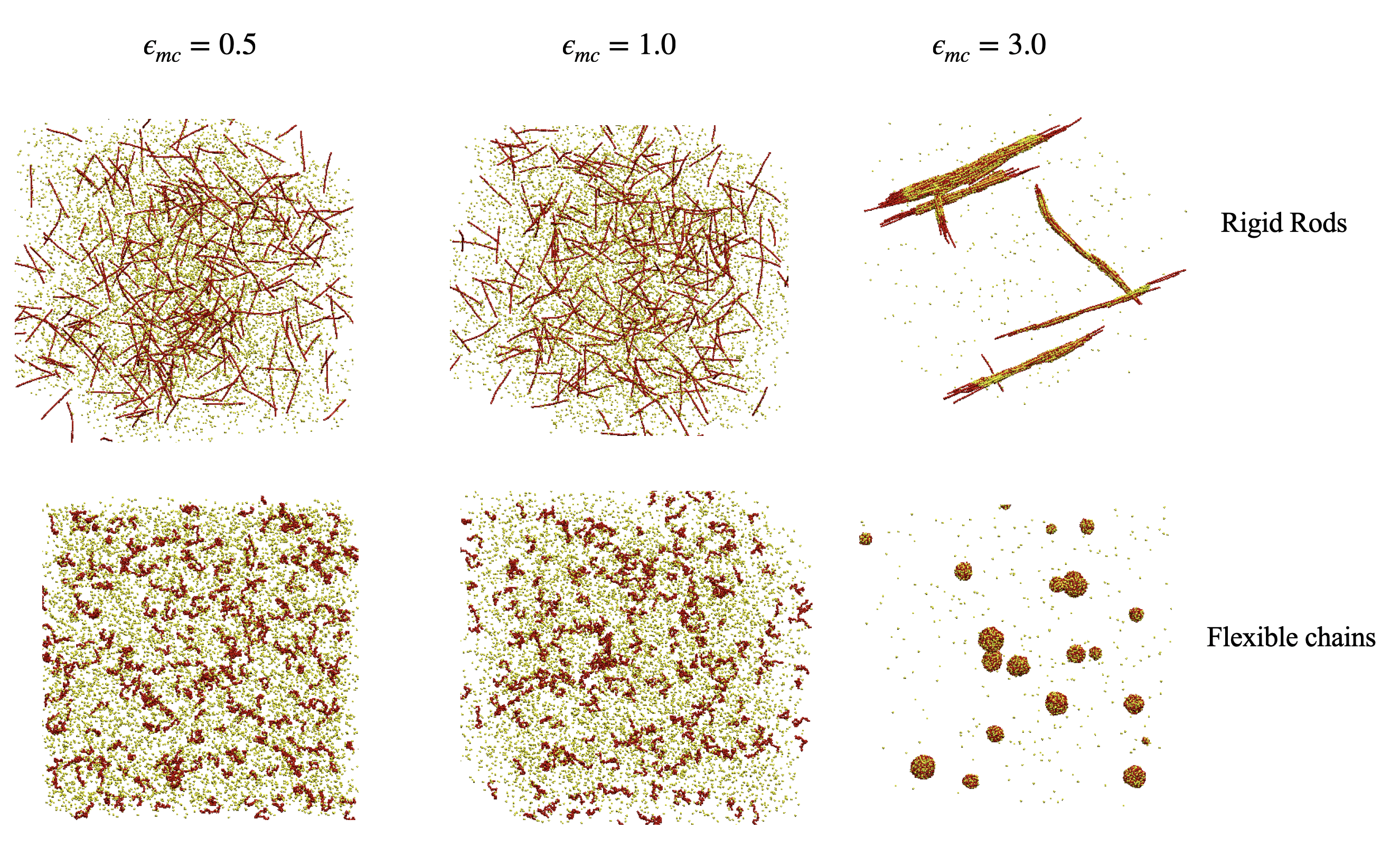}
 \caption{Snapshots illustrating the effect of monomer-crowder attraction strength ($\epsilon_{mc}$) on polymer aggregation. Top row: rigid rod-like polymers form elongated, bundle-like aggregates only at high attraction ($\epsilon_{mc} = 3.0$). Bottom row: flexible polymer chains form compact spherical clusters at much lower attraction strengths. Crowder particles and polymers are shown in red and yellow.}
\label{snapshot_rigid_flexible}
\end{figure}
As described in the Methods section, we consider a system consisting of 300 polymer chains, each composed of 30 monomers, along with 9000 crowders placed within a cubic simulation box of length $L = 200$, resulting in a crowder number density of $\rho_c = 1.125 \times 10^{-3}$. We first investigate whether purely repulsive interactions among the system constituents can induce aggregation at this density. For this purpose, the monomer-monomer, crowder-crowder, and monomer-crowder interactions are modeled using the Weeks-Chandler-Andersen (WCA) potential, which explicitly accounts only for steric repulsion and excludes any attractive component. Figure S1 (Supporting Information) shows the equilibrated snapshot of the system under these purely repulsive conditions, with rigid polymer chains shown in red and crowder particles depicted in yellow. The snapshot reveals a homogeneous spatial distribution of polymer chains and crowder particles throughout the simulation box, without any discernible polymer clustering or aggregation. This outcome clearly indicates that purely steric repulsive interactions alone are insufficient to induce aggregation under the given density and system parameters. In the subsequent sections, we therefore explore the impact of introducing attractive interactions between monomers and crowders, as well as between crowders themselves, while maintaining purely repulsive monomer-monomer interactions, to investigate the emergence and characteristics of polymer aggregation.

Next, we investigate the impact of introducing attractive monomer-crowder interactions ($\epsilon_{mc}$) on the aggregation behavior of neutral polymer chains characterized by inherently repulsive monomer-monomer interactions. At low values of monomer-crowder attraction strength ($\epsilon_{mc}$), polymer chains remain uniformly dispersed throughout the simulation box, irrespective of their flexibility, due to insufficient attractive interactions to overcome entropic and steric constraints. As the attraction strength increases, however, significant aggregation emerges for both rigid rods and flexible polymers. Representative snapshots illustrating this transition for rigid rods and flexible chains at increasing values of $\epsilon_{mc}$ are presented in Figure~\ref{snapshot_rigid_flexible}.  Polymer aggregation is quantitatively characterized by defining two polymer chains as aggregated if any monomer of one chain is within $2\sigma_m$ of any monomer of another chain. Consequently, a polymer chain is considered part of an aggregate of size $m$ if this criterion is satisfied with respect to $m-1$ other chains.

The onset of aggregation is systematically captured by the relative fluctuation ($\chi$) in the pair interaction energy per particle ($E$), defined as:
\begin{equation}
\chi = \frac{\langle E^2 \rangle - \langle E \rangle^2}{\langle E \rangle^2}.
\label{eq:chi}
\end{equation} 
Figure~S2 (Supporting Information) shows the variation of $\chi$ as a function of monomer-crowder attraction strength, $\epsilon_{mc}$. A pronounced peak in $\chi$ signals the onset of aggregation, occurring around $\epsilon_{mc}^* \approx 2.0$ for rigid rods and $\epsilon_{mc}^* \approx 1.7$ for flexible chains. These peaks indicate critical aggregation thresholds, marking structural rearrangements from homogeneous dispersions to finite-sized aggregated structures. 

\begin{figure}
\includegraphics[width=\columnwidth]{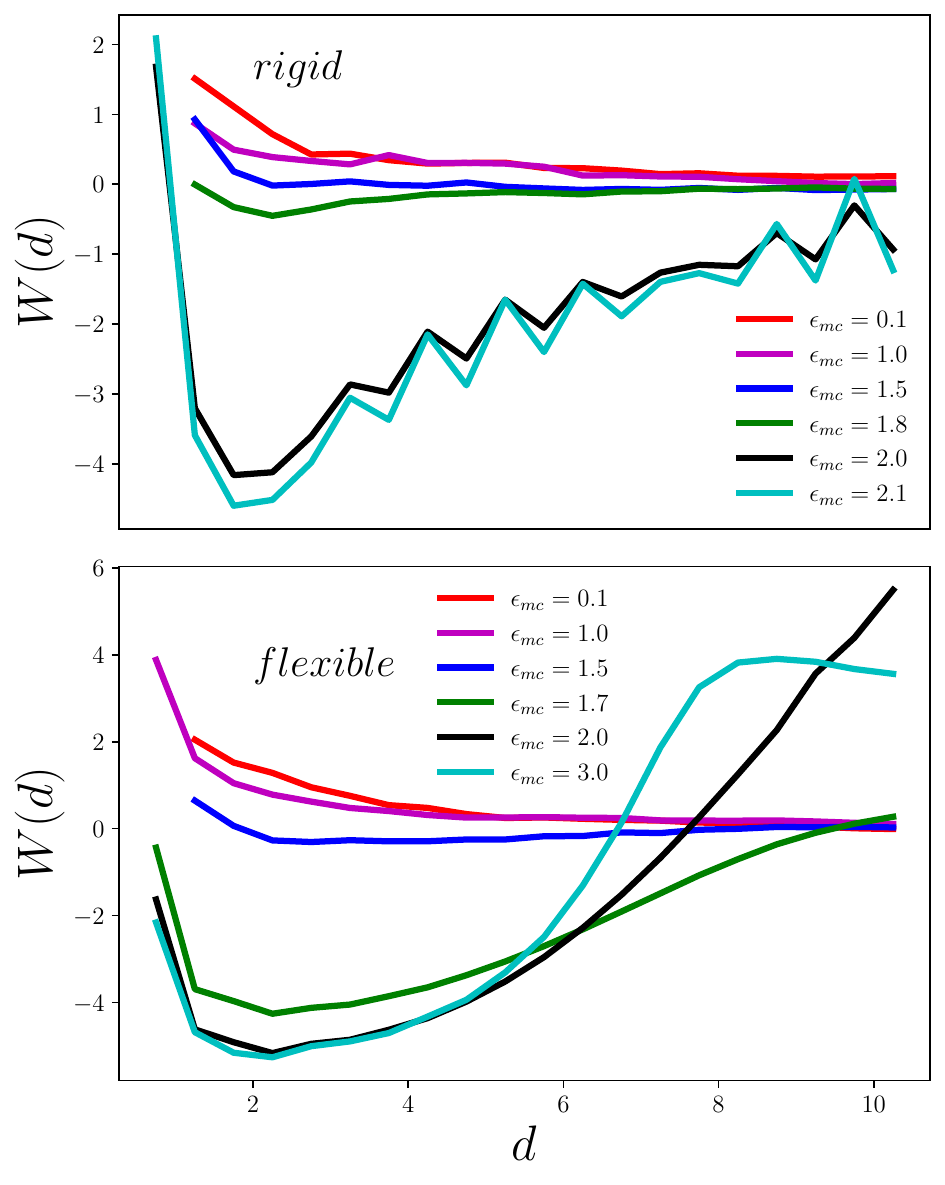}
 \caption{The potential of mean force $W(d)$ as a function of center-of-mass separation $d$ between two polymers, for various monomer-crowder attraction strengths $\epsilon_{mc}$. Top panel: rigid rods. Bottom panel: flexible chains. }
\label{pmf}
\end{figure}

To establish a fundamental understanding of the relationship between single-polymer collapse and multi-polymer aggregation mediated by attractive crowders, we first examine the extended-to-collapsed transition of an isolated flexible polymer chain in an environment mimicking the multi-chain scenario. Specifically, we simulate a single flexible polymer chain consisting of 100 monomers in the presence of 72 crowders, maintaining the same monomer and crowder densities as the multi-chain simulations. Figure S3 (Supplementary Information) shows the radius of gyration, $R_g$, of the single polymer chain as a function of monomer-crowder attraction strength ($\epsilon_{mc}$). A clear transition from an extended to a collapsed conformation is observed, and we extract the critical attraction strength for this transition by fitting the data to a hyperbolic tangent function. The resulting critical value, $\epsilon_{mc}^* \approx 1.6$, closely aligns with the aggregation threshold values observed for multi-polymer systems, particularly for flexible chains ($\epsilon_{mc}^* \approx 1.7$). This agreement strongly suggests that similar underlying physical mechanisms drive both the collapse of individual polymer chains and the aggregation of multiple polymers when mediated by attractive crowders. Physically, this correspondence implies that the monomer-crowder attractions sufficiently strong to induce intrachain bridging (leading to chain collapse) also facilitate interchain bridging (leading to aggregation). To further investigate these interchain bridging interactions explicitly, we calculate the potential of mean force (PMF), $W(d)$, between two polymer chains in the presence of attractive crowders. The PMF provides a direct measure of the effective interaction between polymer chains, mediated by the surrounding crowders, and is derived from the equilibrium pair correlation function ($g(d)$) between two polymer chains separated by distance $d$ using $W(d) = -k_B T \ln g(d)$. In these simulations, we consider two parallel polymer chains (each consisting of 30 monomers) along with 300 crowders at a fixed monomer density of $9.375 \times 10^{-4}$ monomers/$\sigma^3$. Figure~\ref{pmf} shows the calculated PMFs for rigid rods (top panel) and flexible chains (bottom panel) at different values of monomer-crowder attraction strength ($\epsilon_{mc}$).

At low attraction strengths ($\epsilon_{mc} \leq 1.0$), the PMF profiles for both rigid and flexible chains remain predominantly positive or near-zero and monotonically decrease with increasing interchain distance $d$, signifying weak or negligible interchain attraction insufficient to stabilize aggregates. As the attraction strength $\epsilon_{mc}$ increases beyond a critical threshold, however, a pronounced negative minimum emerges in the PMF at short distances, indicating stable attractive interactions strong enough to hold polymer chains together in stable aggregates. Notably, the critical attraction strengths inferred from these PMF minima align remarkably well with the critical aggregation strengths obtained in our multi-chain simulations. Specifically, rigid rods exhibit a clear attractive minimum at approximately $\epsilon_{mc}=2.0$, while flexible chains exhibit a similar feature at approximately $\epsilon_{mc}=1.7$. This close correspondence further supports the notion that the aggregation threshold in many-chain systems fundamentally relates to the onset of crowder-mediated bridging attractions between pairs of polymer chains. An additional insight from Figure~\ref{pmf} is that the depth and width of the attractive well differ significantly between rigid rods and flexible polymers. For rigid rods, the attractive well is deep and well-defined, reflecting strong anisotropic alignment and favorable directional interactions facilitated by bridging crowders. In contrast, the flexible polymer PMFs show shallower, broader minima indicative of weaker, isotropic bridging interactions. This difference is consistent with morphological observations of aggregate structures in multi-chain systems, where rigid rods form large, elongated bundles due to their efficient directional bridging, whereas flexible polymers form smaller, more isotropic clusters due to less efficient, orientation-independent bridging. Thus, the PMF analysis provides a direct microscopic explanation for the observed differences in aggregation kinetics and morphology driven by polymer flexibility. The quantitative consistency across single-chain collapse data, pairwise PMF calculations, and multi-chain aggregation thresholds strongly underscores the utility and importance of understanding minimal subsystems (single polymer collapse and two-chain bridging interactions) to predict complex multi-polymer aggregation behaviors.

The distinct aggregation morphologies observed (Figure~\ref{snapshot_rigid_flexible}) can be rationalized by considering the interplay between polymer rigidity and the bridging effects of crowder particles. In rigid rods, the anisotropic geometry strongly favors aligned bundling, reminiscent of systems involving anisotropic particles such as carbon nanotubes, cellulose nanocrystals, and biological filaments (e.g., actin or microtubules), where alignment and orientation-based aggregation dominate due to directional interactions and geometric constraints~\cite{hunter2020pickering,kalashnikova2011new,dugyala2016nano}. Conversely, the isotropic and flexible nature of polymer chains enables multiple conformations, allowing crowders to effectively bridge distant polymer segments from various directions, thereby forming spherical aggregates with significantly reduced anisotropy~\cite{heyda2013rationalizing,huang2021chain,garg2023conformational}. The differences observed in the maximum cluster size (Figure S4, see (Supplementary Information)) further reflect the fundamental impact of polymer flexibility on aggregation kinetics and morphology. Rigid polymers readily form extended clusters due to their inherent alignment capabilities and strong directional interactions facilitated by crowder bridging. Flexible polymers, however, are limited by conformational entropy and lack inherent alignment preferences, leading to relatively smaller, compact aggregates. The notably larger cluster sizes observed for rigid rods suggest that rigidity not only facilitates directional alignment but also enhances effective attraction between polymer chains via crowder bridging, enabling larger, more stable aggregates to form. These insights align well with prior literature that underscores polymer flexibility as a key determinant in shaping aggregation kinetics and structural outcomes. Studies of rod-like and flexible colloidal particles, proteins, and biopolymers have repeatedly demonstrated how structural rigidity favors extended, aligned bundles, whereas flexibility generally promotes more isotropic, compact assemblies~\cite{dugyala2016nano,kalashnikova2011new,heyda2013rationalizing,brackley2020polymer}. Furthermore, the large disparity in cluster sizes and morphologies between rigid and flexible polymers emphasizes the critical role of anisotropic bridging interactions—mediated by crowders—in modulating aggregation and structural transitions in neutral polymer systems. 

\begin{figure}
\includegraphics[width=\columnwidth]{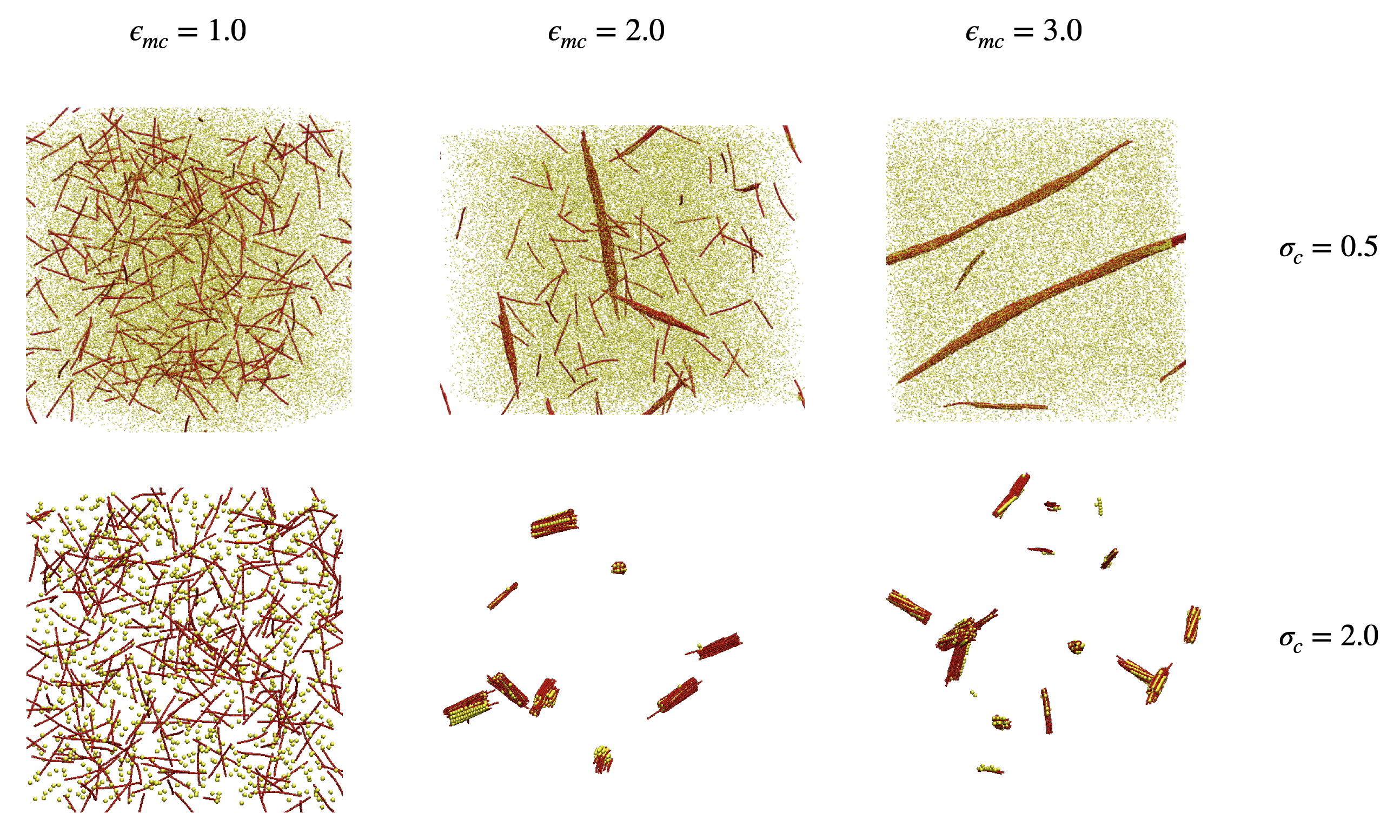}
 \caption{Snapshots of rigid rod-like polymers at different monomer-crowder attraction strengths ($\epsilon_{mc} = 1.0$, $2.0$, $3.0$) for two different crowder sizes: $\sigma_c = 0.5$ (top row) and $\sigma_c = 2.0$ (bottom row). }
\label{scc}
\end{figure}

The size of crowders significantly influences the morphology and the onset of polymer aggregation, revealing distinct aggregation pathways and structural features depending on both crowder size and polymer rigidity. To systematically examine this effect, we considered three different crowder sizes: $\sigma_c = 0.5, 1.0,$ and $2.0$ (Table~\ref{table3}). Snapshots illustrating aggregation morphologies of rigid rods and flexible chains at varying $\epsilon_{mc}$ for crowder sizes $\sigma_c = 0.5$ and $2.0$ are presented in Figures~\ref{scc} and S5(Supplementary Information). For rigid rods (Figure~\ref{scc}), smaller crowders ($\sigma_c = 0.5$) produce larger, more elongated bundles with clear alignment among the rods at intermediate values of monomer-crowder attraction ($\epsilon_{mc}=2.0$), progressively coalescing into very large and oriented clusters at higher attraction strength ($\epsilon_{mc}=3.0$). These elongated structures are indicative of strong anisotropic interactions and effective bridging induced by smaller crowders, which can more readily intercalate between polymers and effectively cross-link them into aligned aggregates. In contrast, larger crowders ($\sigma_c=2.0$) yield compact, fragmented bundles at comparable attraction strengths. The polymers aggregate into short, segmented bundles rather than continuous extended structures. This can be attributed to steric hindrance: large crowders occupy substantial volume and impede the formation of extended, aligned bundles by restricting the accessibility of monomers to bridging interactions. These compact aggregates formed in the presence of large crowders resemble those found in systems where excluded volume dominates, consistent with previous theoretical and experimental studies on depletion interactions, which show that larger crowders effectively enhance osmotic pressure and lead to local clustering rather than extended aggregation \cite{lekkerkerker2011stability,yodh2001entropically,kim2015polymer}.

For flexible polymers, seen in Figure S5 (Supplementary Information), aggregation morphologies distinctly differ from those of rigid rods. At smaller crowder sizes ($\sigma_c=0.5$), flexible chains exhibit well-defined, compact spherical clusters. As the attraction strength increases from $\epsilon_{mc}=1.0$ to $3.0$, these spherical clusters grow in size while maintaining their compact shape. This spherical morphology results from the isotropic nature of flexible polymers, allowing them to easily rearrange and adopt energetically favorable spherical aggregates that minimize surface tension and maximize internal binding interactions. The compact nature of these clusters aligns well with experimental observations of biomolecular condensates and globular aggregates commonly observed in protein-crowder systems, where isotropic polymer-crowder interactions promote liquid-like condensate structures or gel-like aggregates \cite{wang2013phase,zhou2008macromolecular}. When the crowder size is increased to $\sigma_c=2.0$, the aggregates of flexible chains become significantly smaller and fragmented at similar attraction strengths. Large crowders severely restrict polymer chain mobility, thereby limiting aggregation and resulting in numerous small, isolated clusters. This fragmentation can be explained by limited diffusion and reduced bridging capability of large crowders due to their steric bulk. Additionally, large crowders hinder close polymer-polymer proximity, thereby suppressing extensive aggregation and leading to smaller clusters. Similar fragmentation due to steric hindrance and reduced mobility in the presence of large crowders has been reported in simulations and experiments involving polymer-crowder mixtures, particularly in biological scenarios, such as the effect of macromolecular crowding on protein and nucleic acid aggregation within cells \cite{sharp2015analysis,kim2015polymer,jeon2016effects}. The critical attraction strength ($\epsilon_{mc}^*$) required for aggregation clearly decreases as the crowder size increases, as detailed in Table~\ref{table3}. This trend, consistent across rigid and flexible polymers, confirms that larger crowders enhance effective polymer-polymer attraction via stronger depletion interactions and increased osmotic effects, aligning with generalized depletion theory predictions \cite{yodh2001entropically,lekkerkerker2011stability}. Notably, the reduction in critical attraction strength for aggregation parallels the earlier observed single-chain coil-to-globule transitions, providing additional evidence for a unified underlying mechanism governing single-chain collapse and multi-chain aggregation. Finally, we note that achieving complete equilibration in strongly aggregated regimes presents significant challenges due to markedly increased diffusion timescales. Small aggregates formed initially diffuse slowly, requiring increasingly longer intervals to encounter and merge with each other, complicating the study of late-stage aggregation dynamics. These dynamical considerations will be explored in detail in subsequent sections, particularly addressing the impact of monomer-crowder attraction strength on aggregation kinetics.

\begin{table}		
\caption{\label{table3} The critical values of transition $\epsilon_{mc}^*$ for case of coil-globule transition of single flexible chain, aggregation of rigid rod and aggregation of flexible chains at same density for different crowder sizes ($\sigma_c$).}
\begin{ruledtabular}
		\begin{tabular}{lllll}
$\sigma_{c}/\sigma_{m}$  &Single chain &Rigid rods &Flexible chain \\
		\hline
$0.5$ & 1.8 & 2.0 & 1.9 \\
$1.0$ & 1.6 & 2.0 & 1.7 \\
$2.0$ & 1.2 & 1.5 & 1.2 \\

\end{tabular}
\end{ruledtabular}
\end{table}

\subsection{Aggregation Kinetics: Rigid Rods vs. Flexible Chains}
\begin{figure}
\includegraphics[width=0.9\columnwidth]{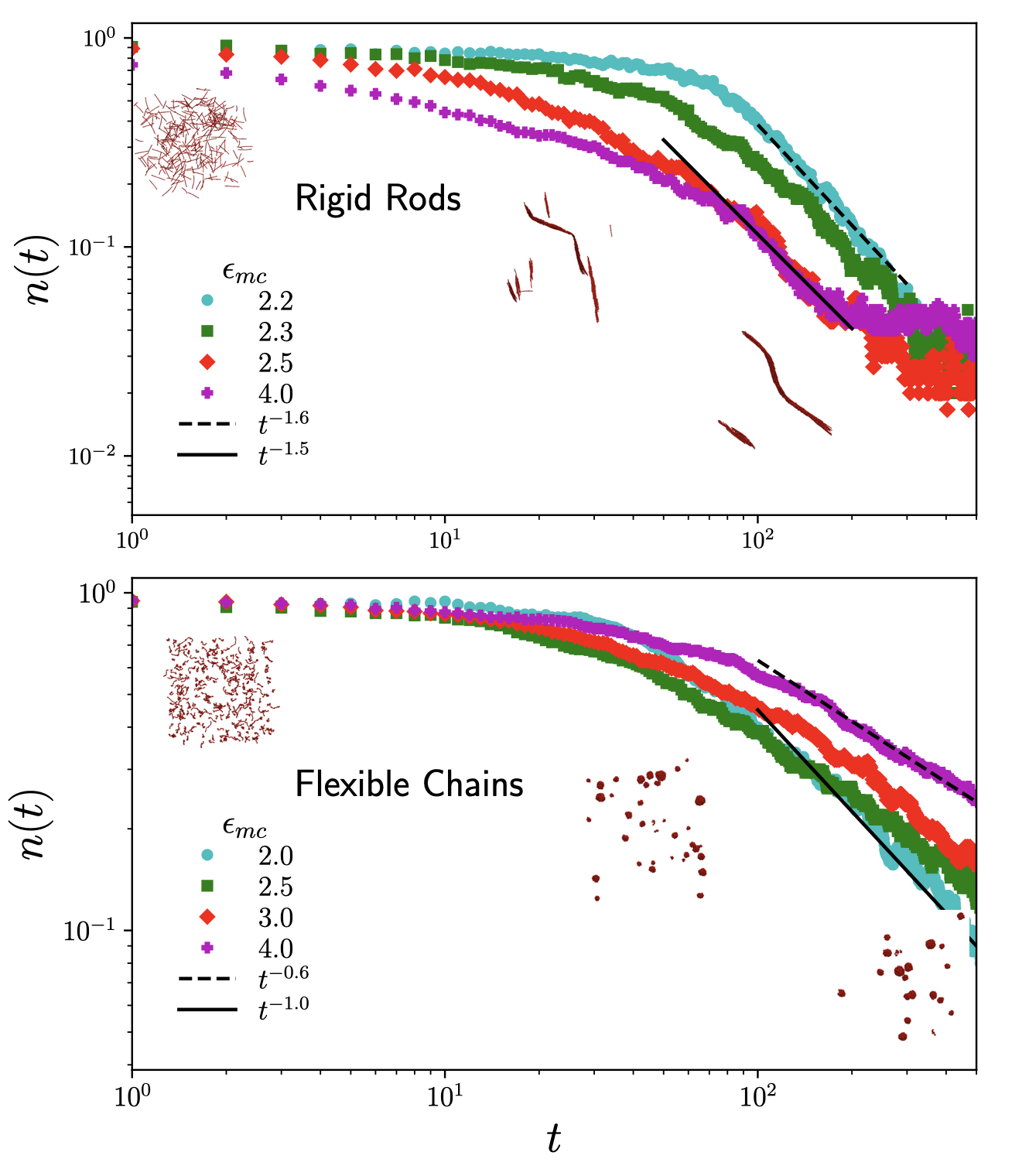}
 \caption{Time evolution of the fraction of unaggregated chains, $n(t)$, for rigid rods (top) and flexible chains (bottom) at various monomer-crowder attraction strengths $\epsilon_{mc}$. The data are log-binned and plotted on log-log axes. }
\label{nwithtime}
\end{figure}

To elucidate the kinetics of polymer aggregation mediated by attractive crowders, we analyze the time evolution of $n(t)$, the average fraction of polymers remaining unaggregated. Figure~\ref{nwithtime} illustrates the temporal evolution of $n(t)$ for rigid rods and flexible polymer chains at different strengths of monomer-crowder attraction ($\epsilon_{mc}$). For clarity, the data is presented in a log-log scale and log-binned to highlight trends across varying time scales, and representative snapshots depicting characteristic aggregate morphologies are included as insets. For both rigid rods and flexible chains, $n(t)$ progressively decreases over time, indicating continued aggregation events. At sufficiently long times, the decay follows a clear power-law relationship: $n(t) \sim t^{-\alpha}$, where $\alpha$ denotes the aggregation kinetics exponent. Notably, we observe significant differences in the decay exponents between rigid and flexible polymers. Rigid rods exhibit a faster aggregation kinetics with a decay exponent of approximately $\alpha \approx 1.6$, whereas flexible chains aggregate more slowly, characterized by $\alpha \approx 1.0$. The pronounced difference in kinetic rates reflects underlying physical distinctions related to polymer structure and flexibility. Rigid rods experience strong directional bridging interactions mediated by crowders, allowing efficient alignment and formation of elongated bundles, thus accelerating the aggregation process. In contrast, flexible chains encounter increased conformational entropy barriers, necessitating substantial local rearrangements before aggregation occurs, leading to slower kinetics.

The dependence of aggregation kinetics on monomer-crowder attraction strength ($\epsilon_{mc}$) is evident for both polymer types, as stronger attractions lead to faster aggregation, reflected by steeper slopes in Figure~\ref{nwithtime}. At lower attraction strengths (e.g., $\epsilon_{mc} = 2.2$ for rigid rods and $\epsilon_{mc} = 2.0$ for flexible chains), aggregation kinetics are relatively slower, highlighting weaker bridging effects. As the attraction strength increases, crowders more effectively mediate polymer-polymer bridging, enhancing aggregation kinetics. The morphological outcomes also differ significantly with polymer flexibility, as observed in the inset snapshots. Rigid rods predominantly form elongated, cylindrical bundles aligned due to anisotropic bridging interactions, whereas flexible polymers assemble into smaller, isotropic, spherical aggregates due to the absence of preferred alignment axes. These observations align with theoretical expectations and simulation studies on anisotropic and isotropic particle aggregation, respectively~\cite{dugyala2016nano,kalashnikova2011new,meakin1983formation,weitz1985limits}.  Furthermore, the power-law exponents observed here can be connected to classical models of aggregation dynamics. The decay exponent of approximately $\alpha \approx 1.6$ observed for rigid rods is consistent with diffusion-limited cluster aggregation (DLCA), suggesting that the aggregation process is primarily determined by cluster diffusion and rapid irreversible sticking upon contact~\cite{meakin1983formation,weitz1985limits}. Conversely, the slower exponent $\alpha \approx 1.0$ for flexible chains is reminiscent of reaction-limited aggregation (RLA), indicative of aggregation processes hindered by energetic or entropic barriers, consistent with previous studies of flexible macromolecules and intrinsically disordered proteins where aggregation is impeded by conformational rearrangements~\cite{heyda2013rationalizing,brackley2020polymer,garg2023conformational}.

Next, we explore how aggregation onset depends on crowder density by systematically varying the crowder number density ($\rho_c$) while keeping the crowder number constant and changing the simulation box size. Specifically, simulations were conducted for box lengths $L = 100, 150, 200, 250$, and $300$, corresponding to crowder densities of $9 \times 10^{-3}, 2.66 \times 10^{-3}, 1.12 \times 10^{-3}, 5.76 \times 10^{-4}$, and $3.33 \times 10^{-4}$, respectively. Figure S6 (Supplementary Information) presents the variation of critical aggregation attraction strength ($\epsilon_{mc}^*$) with crowder density for both rigid rods and flexible polymer chains. The critical value $\epsilon_{mc}^*$ systematically decreases as crowder density increases, underscoring the important role of crowder density in enhancing aggregation by promoting effective bridging interactions. At higher densities, the frequency of monomer-crowder contacts increases, thereby lowering the attraction threshold necessary to induce aggregation. These observations are quantitatively summarized in Table S1 (Supplementary Information), providing critical values of $\epsilon_{mc}^*$ across densities, clearly demonstrating the density dependence for both polymer types. Interestingly, rigid rods consistently require a slightly higher critical attraction strength compared to flexible polymers at corresponding densities. This observation suggests that although rigid rods aggregate faster once aggregation begins (due to alignment and anisotropic bridging), initiating aggregation itself requires overcoming initial directional constraints and spatial orientations. Flexible polymers, owing to their higher conformational adaptability, are more easily influenced by crowders even at lower attraction strengths. These trends are consistent with literature on crowding-driven polymer collapse and aggregation, highlighting how polymer flexibility critically influences both initial aggregation thresholds and the kinetics of subsequent aggregation events~\cite{heyda2013rationalizing,antypov2008computer,huang2021chain}. Our analysis emphasizes the importance of polymer flexibility, crowder-mediated bridging interactions, and crowder density in determining aggregation kinetics and morphology. The rigid rods demonstrate diffusion-limited aggregation behavior characterized by rapid kinetics and anisotropic aligned bundle formation, while flexible polymers exhibit slower reaction-limited aggregation with isotropic, compact cluster formation. 

\subsection{\label{sec:morphology} Morphologies of chains}
\begin{figure} 
\includegraphics[width=0.9\columnwidth]{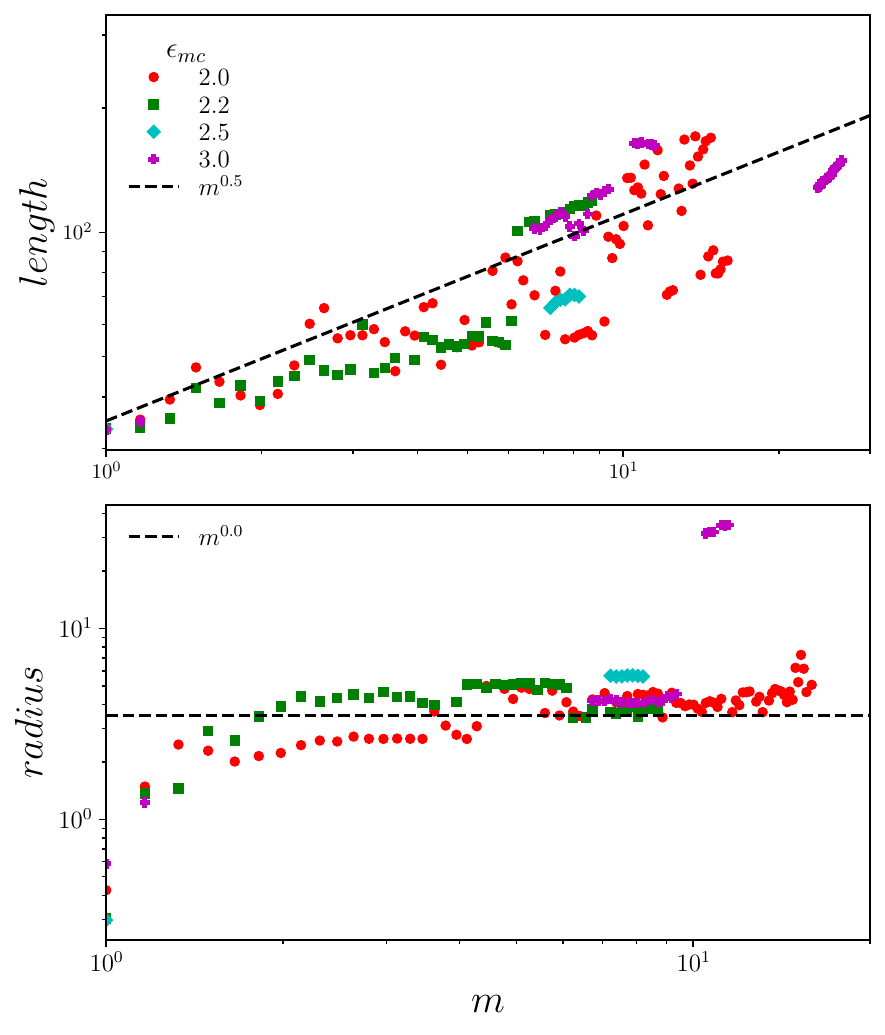}
 \caption{The variation of length and radius of aggregate cylinders with aggregate size for rigid rods at different monomer-crowder attraction strengths $\epsilon_{mc}$. Length scales as $m^{0.5}$ while radius remains nearly constant, indicating linear aggregation growth.} 
 \label{geometry_rigid} 
 \end{figure}
 
\begin{figure}
\includegraphics[width=0.9\columnwidth]{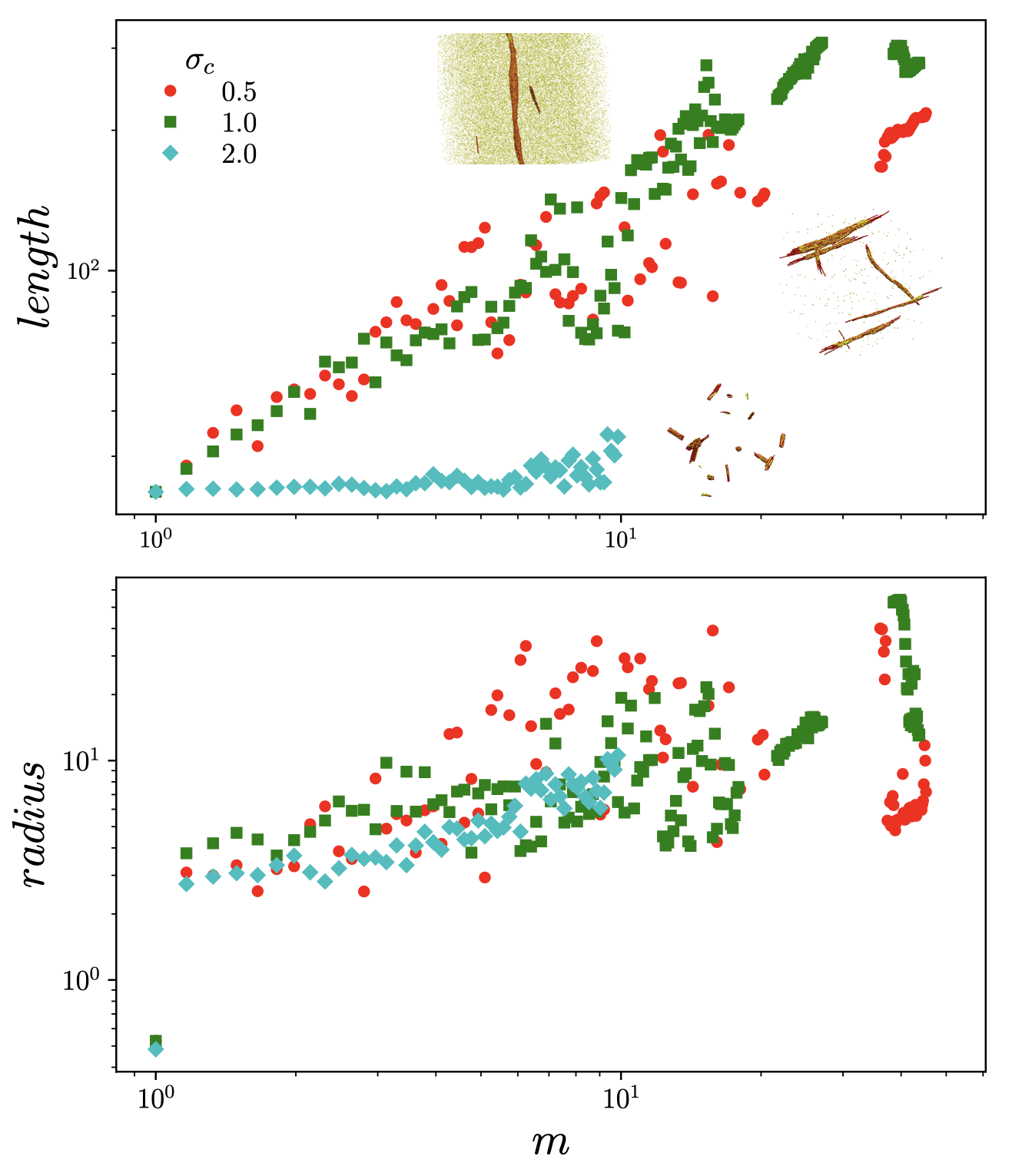} 
\caption{Length and radius of aggregates formed by rigid rods as functions of cluster size for different crowder sizes ($\sigma_{c}$). } 
\label{radiusprolateness_size} 
\end{figure}

 To characterize aggregate morphologies quantitatively, we calculated the eigenvalues of the gyration tensor $S$, whose components are given by: 
 \begin{equation} 
 S_{\alpha \beta}=\frac{1}{N}\sum_{i=1}^{N}r_{i\alpha}r_{i\beta}, \quad \alpha,\beta=1,2,3, 
 \label{eq:gyration} 
 \end{equation} 
 where $r_{i\alpha}$ is the $\alpha^{th}$ component of the position vector $\vec{r_i}$ of particle $i$, measured from the aggregate’s center of mass. The eigenvalues, denoted as $\lambda_1,\lambda_2,\lambda_3$ (with $\lambda_1>\lambda_2>\lambda_3$), allow us to approximate rigid rod aggregates as cylindrical shapes. Figure~\ref{geometry_rigid} demonstrates that aggregate length scales with cluster size $m$ as $m^{0.5}$, while radius remains nearly constant. This scaling indicates a linear or collinear aggregation mechanism, consistent with prior literature on anisotropic systems (carbon nanotubes, actin filaments, and elongated colloidal particles) that similarly form linear bundles due to reduced kinetic barriers for parallel aggregation~\cite{nguyen2002kinetics,bordi2006direct,susoff2008aggregation}. Similar elongated, rod-like morphologies were also observed for rigid charged polyelectrolytes in our earlier study~\cite{tom2017aggregation}. There, multivalent counterions facilitated aggregation by bridging similarly charged polymer backbones. The resulting cylindrical aggregates showed length-dominated growth akin to the neutral rod systems investigated here. However, unlike the neutral crowder-driven bridging, aggregation in charged polymer systems involved additional complexity from electrostatic repulsion between like charges. Despite this difference, the underlying principle of bridging (mediated by counterions or crowders) producing elongated linear aggregates remains common between the two systems, highlighting the universality of bridging-induced anisotropic growth. Crowder size significantly modulates morphology, as demonstrated in Figure~\ref{radiusprolateness_size}. For small crowders ($\sigma_c=0.5,1.0$), length and radius both modestly increase, suggesting efficient, uniform bridging along rods. Conversely, larger crowders ($\sigma_c=2.0$) predominantly increase aggregate radius while maintaining near-constant length. This indicates a transition in bridging mechanism: smaller crowders bridge rods linearly (collinear aggregation), while larger crowders induce side-by-side aggregation due to stronger localized attractions. This crowder-size-dependent bridging mechanism parallels the role of counterion valency in charged polymer systems, where higher valency counterions similarly facilitate stronger, more localized bridging, yielding thicker and shorter aggregates~\cite{tom2017aggregation,varghese2012aggregation}.

\begin{figure} 
\includegraphics[width=0.9\columnwidth]{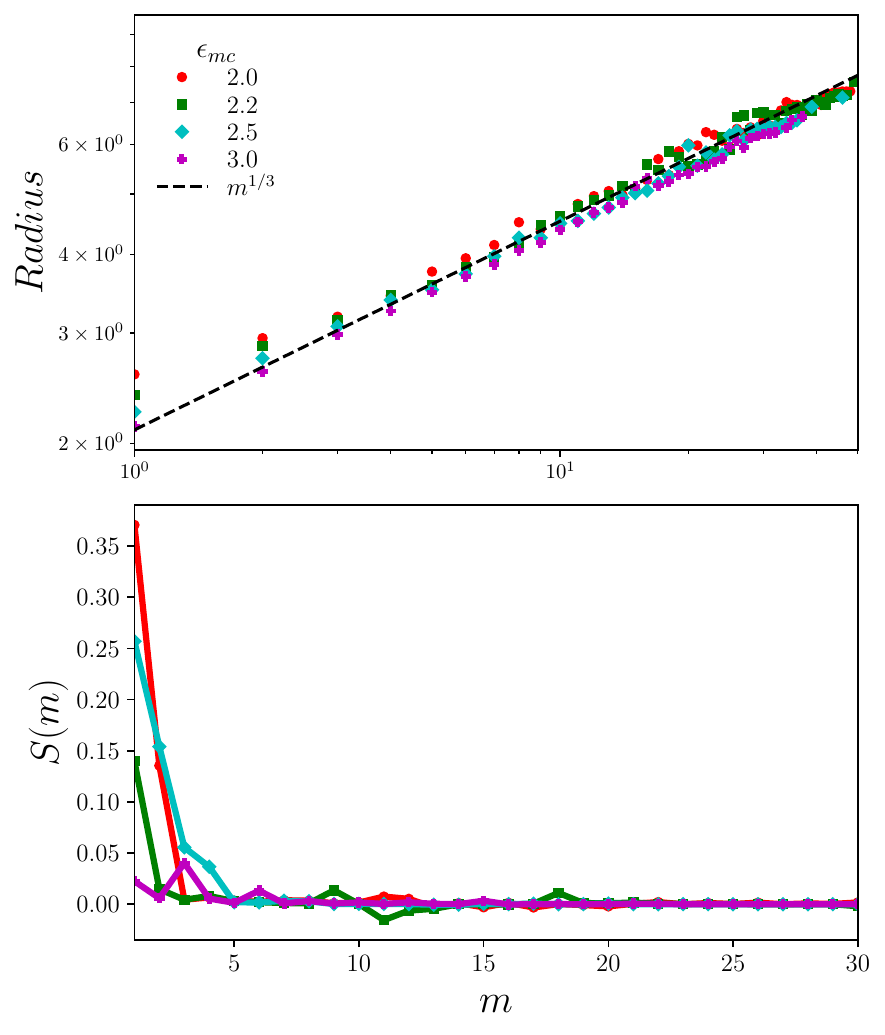} 
\caption{(a) Radius and (b) prolateness of aggregates for flexible polymers as functions of aggregate size at different $\epsilon_{mc}$.} 
\label{radiusprolateness_flexible} 
\end{figure}

For flexible polymer aggregates, we defined the radius as $R_g = \sqrt{\lambda_1+\lambda_2+\lambda_3}$, which exhibited a scaling of $m^{1/3}$ (Figure~\ref{radiusprolateness_flexible}(a)). This scaling confirms compact, spherical aggregates typical of isotropic, diffusion-limited aggregation processes. To further quantify shape, we computed prolateness $S(m)$: \begin{equation} S(m)=\frac{\prod_{i=1}^{3}\lambda_i(m)-\bar{\lambda}_m^3}{\bar{\lambda}_m^3}, \label{eq:prolateness} \end{equation} where negative values indicate oblate shapes, positive values denote prolate shapes, and zero corresponds to perfect spheres~\cite{dima2004asymmetry}. Figure~\ref{radiusprolateness_flexible}(b) illustrates that aggregates begin as mildly prolate ellipsoids ($S>0$) and approach spherical symmetry ($S\approx 0$) as aggregation progresses. Figure S7 (Supporting Information) shows the variation in the scaled radius of gyration ($R_g^1(m)/R_g^1(1)$) of a single flexible polymer chain within an aggregate of size $m$, plotted as a function of $m$ for different monomer-crowder interaction strengths ($\epsilon_{mc}$). For all considered values of $\epsilon_{mc}$, the scaled radius increases as the aggregate size $m$ grows, indicating that chains within larger aggregates become more extended relative to their conformation in isolation. This implies that as the aggregate size increases, individual flexible chains are forced to stretch out more significantly within clusters due to geometric constraints and entropic repulsion from neighboring chains, a phenomenon observed charge polymer aggregates~\cite{tom2017aggregation}. Moreover, the stronger the monomer-crowder attraction (higher $\epsilon_{mc}$), the more significant the chain extension within aggregates. This is because higher attraction values result in tighter aggregates, requiring individual chains to adopt more extended configurations to minimize free energy.

Our previous investigation of charged flexible polyelectrolytes also found compact, near-spherical morphologies~\cite{tom2017aggregation}. There, multivalent counterions mediated short-range attractions, facilitating entropic gains from counterion release upon aggregation. Both systems share this compact spherical morphology, underscoring that short-range bridging attractions, regardless of whether they originate from crowders (neutral system) or counterions (charged system), generally drive flexible polymer aggregation toward compact isotropic aggregates. However, the charged polymer aggregates exhibited significant internal rearrangements and fragmentation events driven by polymer entanglement and reversible electrostatic bridging, leading to slower kinetics and complex aggregate interiors. In contrast, neutral flexible polymers with crowders aggregate irreversibly, forming simpler internal structures without frequent fragmentation, reflecting differences in the reversibility and nature of bridging.
Moreover, crowder size in neutral systems plays a role similar to counterion valency in charged polymer systems. Larger crowders mimic higher valency counterions by introducing stronger localized bridging attractions, lowering the critical attraction strength needed for aggregation (Table~\ref{table3}), and producing smaller, more compact aggregates. This analogy helps rationalize aggregation morphologies observed across these different systems by highlighting the universal importance of bridging strength and spatial scale in governing aggregate shape and size. Thus, despite fundamental differences in the underlying microscopic interactions (electrostatic in charged polymers, entropic/excluded-volume in neutral polymers), both systems manifest similar overarching morphological principles driven by bridging-induced short-range attraction. The flexibility-dependent transition from elongated linear morphologies (rigid) to compact spherical aggregates (flexible) underscores the universality of aggregation mechanisms modulated by bridging agents, whether neutral crowders or charged multivalent counterions.

\section{\label{sec:summary} Discussion and Conclusions}

In this study, we used molecular dynamics simulations to investigate whether attractive crowders can induce clustering in neutral rigid and flexible polymers. We observed that when particles interact purely through Weeks-Chandler-Andersen (WCA) interactions, aggregation of polymer chains does not occur at low density. This indicates that depletion interactions alone, at these densities, are insufficient to drive polymer clustering. However, when crowders exhibit attractive interactions with polymer monomers, clustering emerges above a critical threshold value of the monomer-crowder attraction parameter, $\epsilon_{mc}^*$. The critical aggregation value $\epsilon_{mc}^*$ is influenced by factors such as crowder density, crowder size, and the bending rigidity of polymer chains. Increasing the crowder density lowers the critical attraction strength $\epsilon_{mc}^*$, facilitating aggregation at lower attraction strengths, likely due to enhanced effective osmotic pressure exerted by the crowders, consistent with classical depletion studies by Asakura and Oosawa~\cite{asakura1954interaction}. Unlike purely depletion-driven aggregation, which arises only at sufficiently high crowder volume fractions due to entropic forces, our simulations clearly demonstrate aggregation at moderate crowder densities only when an explicit crowder-polymer attraction exists (bridging interactions). Thus, bridging dramatically lowers the aggregation threshold compared to depletion mechanisms, highlighting the essential role of specific crowder-polymer affinity.

Our simulations further revealed substantial differences in aggregation kinetics between rigid rods and flexible chains. For rigid rods, aggregation kinetics depend explicitly on the attraction parameter $\epsilon_{mc}$, whereas for flexible chains, aggregation kinetics are largely independent of $\epsilon_{mc}$. Rigid rods can adopt multiple alignments (orthogonal and parallel), similar to behaviors previously reported in charged polymer systems, where multivalent counterions mediate directional bridging interactions~\cite{tom2016aggregation,tom2017aggregation}. In contrast, flexible chains tend to collapse into a single, more isotropic conformation due to their high conformational entropy, limiting their sensitivity to varying attraction strength. The role of crowder size was also investigated for both rigid rods and flexible chains. As crowder size ($\sigma_c$) increased, the critical attraction strength $\epsilon_{mc}^*$ decreased, indicating larger crowders promote aggregation at lower attraction strengths. This is consistent with theoretical predictions on depletion interactions, where larger crowders induce stronger osmotic pressures and bridging interactions, leading to polymer compaction at lower attraction strengths~\cite{yodh2001entropically}. Kim et al.~\cite{kim2015polymer} observed similar trends in polymer confinement studies, concluding larger crowders reduce available configurational space more effectively, enhancing polymer compaction.

Morphological analysis indicated clear dependencies on polymer flexibility and crowder size. Rigid rods formed elongated cylindrical bundles, with aggregate length scaling as $m^{0.5}$ while the radius remained nearly constant, suggesting a predominantly collinear aggregation mechanism. This mechanism mirrors charged polymer aggregation behavior, where multivalent counterions similarly induce linear bundling by bridging charged polymers, also yielding a length scaling exponent near $m^{0.5}$~\cite{tom2017aggregation,varghese2012aggregation}. This highlights a universal role for bridging interactions, whether mediated by attractive neutral crowders or charged counterions, in facilitating anisotropic aggregation. Interestingly, crowder size in neutral polymer systems plays a role analogous to counterion valency in charged polymer systems. Larger crowders strengthen local bridging interactions, promoting thicker, more compact aggregates. Similarly, increasing counterion valency in charged polymer systems produces shorter, thicker aggregates due to enhanced electrostatic bridging~\cite{tom2017aggregation,varghese2012aggregation}. These parallels underscore the universality of aggregation mechanisms driven by bridging interactions.

Flexible polymer aggregates formed compact, spherical clusters with radius scaling as $m^{1/3}$, characteristic of isotropic diffusion-limited aggregation. This behavior resembles charged flexible polymers, where multivalent counterions also drive compact aggregation due to counterion release and entropic gains~\cite{tom2017aggregation}. However, charged polymer aggregates often exhibited reversible rearrangements and fragmentation events, whereas neutral flexible polymer systems formed simpler, irreversibly aggregated structures, underscoring differences in bridging interaction reversibility and internal cluster dynamics. Further, we observed significant chain extension within flexible polymer aggregates as aggregate size increased, reflecting geometrical constraints and entropic repulsions. Similar internal extensions have been noted in charged polymer aggregates, driven by compact packing and bridging interactions~\cite{tom2017aggregation}. Contextualizing our findings, Heyda et al.~\cite{heyda2013rationalizing} explained bridging-induced attraction between macromolecular surfaces via crowder-mediated interactions using effective one-component and Flory–de Gennes mean-field models. Sapir et al.~\cite{sapir2016macromolecular} further validated bridging concepts by demonstrating free energy minima that favor bridging configurations. Our findings align closely with these theoretical insights, reinforcing bridging interaction concepts across diverse polymer systems.

Additionally, single flexible polymer chains exhibited coil-to-globule transitions at $\epsilon_{mc}^*$ values nearly identical to those driving multi-chain aggregation. This indicates that similar bridging and depletion interactions underlie single-chain collapse and multi-chain clustering, highlighting universal aggregation mechanisms across polymer scales. In summary, our simulations demonstrate that attractive crowders effectively drive aggregation in neutral polymer systems, lowering the energetic threshold for polymer clustering in a manner analogous—though distinct at microscopic scales—to charged polymer systems mediated by multivalent counterions. Aggregation behavior depends sensitively on crowder density, crowder size, and polymer rigidity, with larger and more attractive crowders promoting aggregation at lower attraction strengths. Insights from this study inform our understanding of polymer aggregation phenomena across synthetic and biological contexts, where crowding agents such as proteins, osmolytes, and macromolecules similarly modulate polymer organization, phase separation, and dynamics. Our insights have relevance to biological systems, such as cytoskeletal networks where bridging proteins modulate filament bundling, and synthetic formulations, where polymer aggregation must be precisely controlled. For instance, the aggregation thresholds and morphologies we identify provide clear strategies to either enhance stability (by avoiding bridging conditions) or promote controlled aggregation (via tuning crowder-polymer affinity).

\begin{acknowledgments}
SV is grateful for critical discussions with R Rajesh prior to this work. The simulations were carried out on the high performance computing machines Nandadevi and Kamet at the Institute of Mathematical Sciences.
\end{acknowledgments}

\bibliography{aggregation}
\end{document}